# Human Activity Recognition from Knee Angle Using Machine Learning Techniques

Farhad Nazari*, *Member*, IEEE, Darius Nahavandi, *Member*, IEEE, Navid Mohajer, and Abbas Khosravi, *Senior Member,* IEEE

Institute for Intelligent System Research and Innovation (IISRI), Deakin University, Australia.

*Abstract—* Human Activity Recognition (HAR) is a crucial technology for many applications such as smart homes, surveillance, human assistance and health care. This technology utilises pattern recognition and can contribute to the development of human-in-the-loop control of different systems such as orthoses and exoskeletons. The majority of reported studies use a small dataset collected from an experiment for a specific purpose. The downsides of this approach include: 1) it is hard to generalise the outcome to different people with different biomechanical characteristics and health conditions, and 2) it cannot be implemented in applications other than the original experiment. To address these deficiencies, the current study investigates using a publicly available dataset collected for pathology diagnosis purposes to train Machine Learning (ML) algorithms. A dataset containing knee motion of participants performing different exercises has been used to classify human activity. The algorithms used in this study are Gaussian Naive Bayes, Decision Tree, Random Forest, K-Nearest Neighbors Vote, Support Vector Machine and Gradient Boosting. Furthermore, two training approaches are compared to raw data (de-noised) and manually extracted features. The results show up to 0.94 performance of the Area Under the ROC Curve (AUC) metric for 11-fold cross-validation for Gradient Boosting algorithm using raw data. This outcome reflects the validity and potential use of the proposed approach for this type of dataset.

## I. Introduction

Human Activity Recognition (HAR) is a technology to classify the motion of the human according to the input from one or multiple sources based on pattern recognition [1]. It has a crucial role in different applications ranging from monitoring and security systems to controlling human-in-the-loop systems [2]. HAR can be potentially a (HRI) [3] in human assistive devices such as orthoses and exoskeletons. Despite so many challenges, HAR has been considered an exciting and popular research topic [4].

HAR aims to recognise different activities like sitting, standing, walking, and stair climbing based on the input from wearable sensors such as inertial sensors [5], i.e. IMUs [6] or external sensors like motion sensors [7], cameras [8] and depth sensors [9]. This monitoring has many use cases, from smart homes [10] and security [11] to elderly care [12], health care [13] and education system [14]. This technology can be used to promote a healthy lifestyle [15], preventing unhealthy habits [16], fall detection [17], and condition tracking [18].

In the case of human assistive devices [19], this classification can be used as the input to the control system to adapt to the assistance ratio [20] of the active orthosis or exoskeleton based on the activity type and intensity of the performing task. Alternatively, it can be used as a redundant system to check whether the robot's assistance is aligned with the user's intention or not. HAR can also be implemented in the safety system to determine if the user is moving towards a hazard or losing stability and act accordingly [21]. In these cases, human motion is either a gesture or activity [22]. A gesture is the movement of hands, which conveys a message, either to another person or a machine. On the other hand, an activity is the general movement of the body, like walking, running, playing tennis, etc.

The study of human activity recognition goes back to the 1990s when Foerster et al. had first tried to detect body posture and motion based on accelerometry [23]. They found that despite discrepancies regarding particular participant movements, the overall correlation between the kinematic analysis and observed action was acceptable. Fallmann and Kropf used the continuous Hidden Markov Model (cHMM) to recognise human activity from a gyrometer and an accelerometer with 97% and 73% precision and recall, respectively [24]. They concluded that the gyrometer has no contribution to classification, and the effect of filters was insignificant. Kim et al. presented a video-based HAR using depth maps and skeleton joints features [25]. In their study, a depth map tracks human skeletal profile, producing 23 skeletal joints. Then, joint distances and centroid point and magnitude are fed into the hidden Markov model, resulting in 84% classification accuracy among nine activities.

One commercially successful example of video-based HAR technology is the Microsoft® Kinect console [26]. Utilising a colour camera and depth sensor, it can extract skeleton joint positions and user's orientation [22]. This feature will be fed into a clustering algorithm to extract a set of distinctive feature vectors known as "codebook", which will be used to classify human motion [27].

Numerous studies in the literature addressed HAR and pattern recognition. However, most researchers designed a precise experiment for data collection to train their pattern recognition model. The limitations of this approach are twofold. Firstly, due to the expensive and time-consuming nature of data collection, the volume of the collected data is almost always limited. This undermines the generalisation ability of the model to every person. Secondly, the final models are only helpful in recognising activities predefined in the data collection process, which restricts the model's versatility as it might not apply to other use cases in the future. The vast amount of datasets collected by experts and researchers worldwide known as big data could help develop more versatile models with the ability to generalise to more people with different characteristics. The limitation here is that these datasets have been collected for other purposes and might not be helpful for initially targeted applications.

This study aims to test the idea of using data that have been initially collected for other purposes – in this case, pathology diagnoses – for human activity recognition. This evaluates the potential of using already existing data to achieve better results in HAR. This research also investigates the possibility of recognising human activity from the movements in just one joint, the knee. The algorithms used in this study are Gaussian Naive Bayes (NB), Decision Tree (DT), Random Forest (RF), K-Nearest Neighbors Vote (KNN), Support Vector Machine (SVM) and Gradient Boosting (GB). Furthermore, two training approaches are compared to raw (de-noised) data and manually extracted features.

This paper is organised as the following. The next section introduces the selected dataset as well as the processing techniques. Section III will be dedicated to the classification methodology and solution approach. Section IV shows the results of the classification for the selected methods. And finally, the concluding remarks and proposed future works will be discussed in Section V.

## II. Data Analysis

This section introduces the selected dataset and discusses data processing and feature extraction algorithms in detail.

### A. Dataset

The UCIEMG dataset in Lower Limb Data Set collected by Sanchez et al. [28] to address the pathological abnormalities [29] is used in this study. It includes data from 22 male subjects performing three different exercises, 11 of which have some knee abnormality. The exercises involve performing knee exercises in sitting and standing positions and normal walking. The knee motion data were collected using a goniometer with the sample rate of 50Hz upsampled to 1000Hz.

### B. Data preparation

In the applied dataset, to minimise the effect of health problems on the models, data corresponding to participants with knee abnormalities were ignored. What remains is the time-series data of knee angle movement from 11 participants without any diagnosed abnormality corresponding to three different activities. Furthermore, to minimise the effect of unwanted noise, the signals are filtered using a low pass filter with an upper band of 20Hz. Two-sided filters have been used

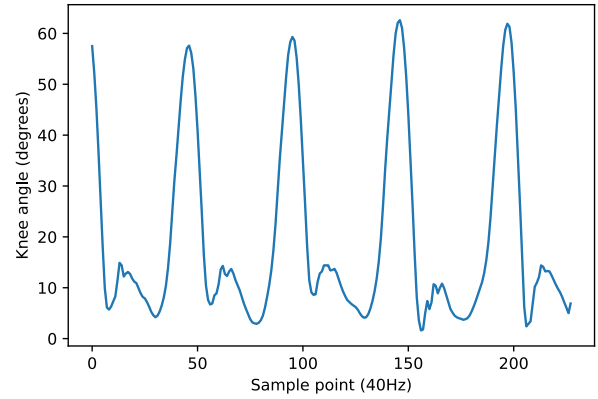
Fig. 1. Recorded knee angle for march over time.

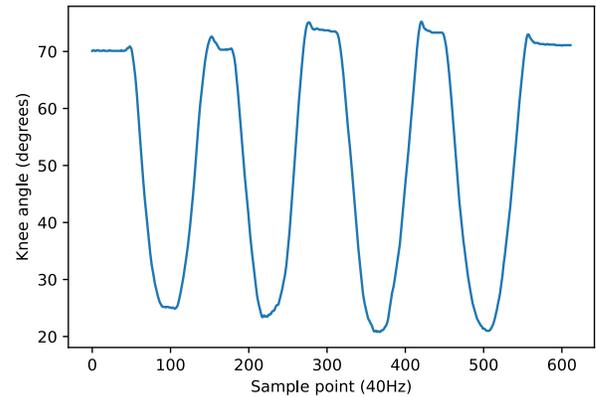
Fig. 2. Recorded knee angle for extension of the leg from sit position over time.

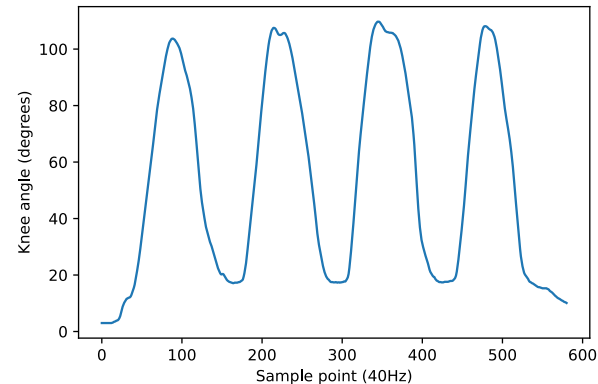
Fig. 3. Recorded knee angle for flexion of the knee stand position over time.

to neutralise the latency effect of the low-pass filter. Finally, to decrease the size of the data and consequently computation cost, the signals were downsampled to 40Hz, which seems enough to preserve the character of the signals.

Figures 1 - 3 show the knee angle position for each exercise over time. The x-axis shows the sample position of the data point downsampled to 40Hz, and the y-axis shows the knee angle position at that point. These figures show that the motion cycle time for different exercises is not the same, making it difficult to choose an appropriate moving window size for the classification task [30]. Especially the gait exercise seems to be more than three times faster than two other activities. We chose a window size of 80 samples in this study, equivalent to two seconds of activity. This corresponds to roughly half a cycle of sitting and standing exercise and a bit less than two cycles of gait exercise. Another approach here could be to compress the data to roughly the same feature size [31] or activity cycle time. Despite being more efficient in training ML algorithms, the second approach might not be the best practice for real-life applications where we do not necessarily know the activity cycle time.

*C. Feature extraction*

Two different approaches for training the Machine Learning (ML) algorithms are compared with each other. The first one is to directly feed the window of preprocessed data to the algorithm as the input [32]. The second approach is to manually extract some features from the time series data and use them as the input to the algorithm [33]. The labels here are the type of activity.

Six features have been extracted for this study: minimum and maximum values, mean (M), median (Med), standard deviation (σ) and mean absolute deviation (MAD). Equations one to four show the calculation of features:

$$M = \frac{1}{n}\sum_{i=1}^{n} x_i \quad (1)$$

$$Med(X) = X\left(\frac{n}{2}\right) \quad (2)$$

$$\sigma = \sqrt{\frac{\sum_{i=1}^{n}(x_i - \bar{x})^2}{n}} \quad (3)$$

$$MAD = \frac{1}{n}\sum_{i=1}^{n} |x_i - \bar{x}| \quad (4)$$

where:
- n is the number of values in each window,
- $x_i$ is the dataset value,
- X is the ordered list of values in the data set, and
- $\bar{x}$ is the population mean.

### III. CLASSIFICATION AND TUNING HYPERPARAMETERS

To evaluate the possibility of human activity recognition from one input, it is aimed at training different ML algorithms on any random two-second portion of knee activity data. The algorithms used here - as mentioned earlier - are NB, GB, DT, RF, KNN and SVM.

The input features are either the raw data of the entire moving window of the time series or the extracted features from that window in each step. As these windows overlap in the data corresponding to every activity of the participants, in order to prevent the data leakage from the training data to the test data, it was split based on the subject before model training. The k-fold cross-validation method has been used to train and test algorithms on different combinations of tests and train subjects where K equals 11.

Hyperparameters are those values used to control the learning process of the ML algorithm. Some popular methodologies for hyperparameter optimisation are Grid Search [34], Random search [35], Bayesian [36], Gradient-based [37], Evolutionary [38], and Population-based [39] optimisations. The current study uses the Grid Search method to tune the hyperparameters of each ML algorithm. For this purpose, data from subjects 1 to 10 have been fitted to each algorithm and scored on the 11[th] subject. Then, the best parameters are extracted to use in the training of the algorithms for 11-fold cross-validation.

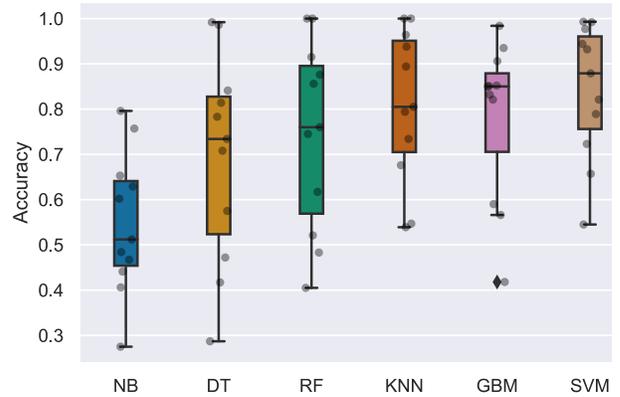

Fig. 3. Cross-validation performance accuracy of each algorithm using the raw data.

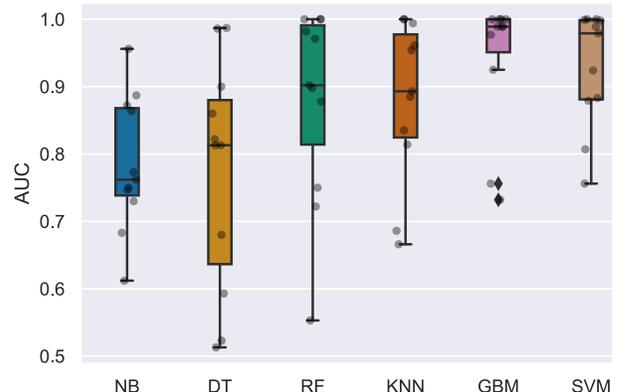

Fig. 4. Cross-validation performance AUC of each algorithm using the raw data.

### IV. CLASSIFICATION RESULTS

In this section, the results of the classification for the aforementioned approaches are shown and discussed.

*A. Classification of raw data*

The highest accuracy was for the Support Vector Machine model, with an average of 84.1%. The lowest accuracy, on the other hand, was for the Naive Bayes model with 54.7%. Decision Tree, Random Forest, K-Nearest Neighbors and Gradient Boosting had an average accuracy of 69.2, 74.3, 80.8, and 78%, respectively. Figure 3 shows the 11-fold cross-validation performance accuracy of each algorithm using raw data.

As we use less than 10% of our data for testing in each step, accuracy might not be the best measure for the algorithm's performance. Looking at the Area Under the ROC Curve (AUC) in Figure 4, we observe less variance and more consistency over different subjects than accuracy. The highest average performance is 0.942 for the GB model, and the lowest is 0.772 for the DT one. NB, RF,

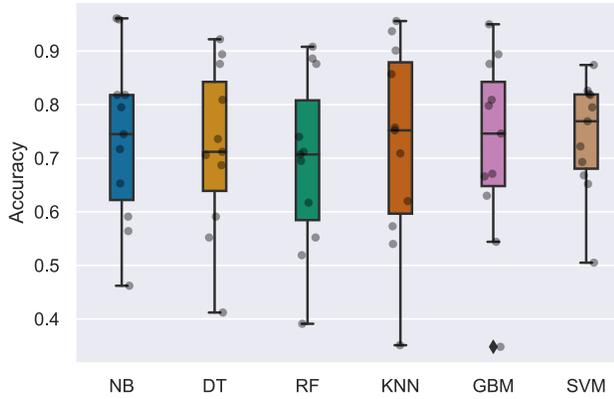
Fig. 5. Cross-validation performance accuracy of each algorithm using extracted features.

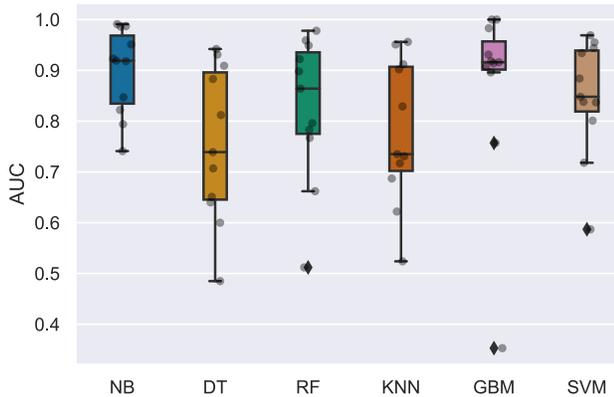
Fig. 6. Cross-validation of AUC for each algorithm using extracted features.

KNN and SVM had average AUC of 0.785, 0.878, 0.881, and 0.929, respectively.

### B. Classification on extracted features

By using the extracted features as input to our ML algorithms, the accuracy increased by 25.6% and 3.6% for NB and DT models and decreased by 7.5, 11.8, 8.5 and 13.6% for RF, K-NN, GB, and SVM, averaging 73.5, 71.8, 69.1, 72.3, 71.9, and 74%, respectively. Figure 5 shows the 11-fold cross-validation performance accuracy of each algorithm using extracted features.

The situation for AUC is a bit different where it rose by 13.1% and 1.9% for DT and KNN and fell by 3.9, 11.3, 12.3 and 8.8% for NB, RF, GB, and SVM, averaging at 0.754, 0.779, 0.898, 0.826, 0.873 and 0.847, respectively. Figure 6 shows the 11-fold cross-validation of AUC for each algorithm using extracted features.

Overall, on average, the performance of ML classification for human activity recognition on extracted features decreased by 2.0 and 3.5% based on accuracy and AUC measures compared to using the raw data.

### C. Computation cost

One of the motivations behind extracting a handful of features for ML models is to reduce computation cost by lowering the dimensionality of the input data. We predicted that the processing time for training each algorithm would drop significantly as we reduce the dimensionality of the features from 80 to just six inputs. For this purpose, we recorded the time just before defining the model and after finishing training to calculate the processing time.

All the computations here have been done on a 10th Generation Intel® Core™ i7 10750H Processor on a Dell laptop with 16GB of RAM, in Visual Studio code editor operating in Microsoft Windows 10 pro. It should be noted that the CPU was throttling around 4.08GHz for single-threaded applications and 3.2GHz for multi-threaded ones.

As shown in Table 1, for NB, DT, RF and GB, the average training time decreased by 82.2, 78.3, 81.8 and 91.6%, respectively. However, the dramatic increase in training time of the KNN model was not aligned with our predictions. This cannot be justified by the difference in hyperparameters, as the best parameters of Grid Search for raw data and extracted features for KNN were identical. Also, this is contrary to the known "Dimensionality Curse" [40] that happen to KNN with increasing dimensionality. The training time of the SVM model increased by 10.3% on the extracted features compared to raw data. This was predictable as SVM tend to perform better on higher dimensions. In other words, the raw data is more separable for the SVM model.

When it comes to testing time, the speed difference for DT, RF and GB for raw data and extracted features is negligible, whereas the performance cost for NB, KNN and SVM decrease by 41.4, 91.8 and 59.4% when switching from raw data to extracted features. Despite being fast in training, the KNN model performs slowly in prediction, especially on raw data, making it an expensive model to run. Similarly, SVM takes a lot of time on both extracted features and raw data. On the other hand, GB performs reasonably fast on both extracted features and raw data, making it expensive to train but reasonable to run.

## V. Conclusion and Future Work

Human Activity Recognition (HAR) is a practical solution for various applications such as human assistance. Pattern recognition is the centrepiece of HAR, and it can enable human-in-the-loop control of different systems such as orthoses and exoskeletons. Most reported studies use a small dataset collected from an experiment for a specific purpose. This approach suffers from a lack of generalisation of the outcome to different people with different biomechanical characteristics and health conditions. Additionally, it fails to be implemented for applications other than the original experiment.

This study investigated the recognition of human activity using the input from knee movement. One channel (knee position) of normal participants from a dataset originally collected for pathology diagnoses were used. Two-second windows of activity were selected as the input to Machine Learning (ML) algorithms in order to classify the activity. The results exhibited up to 84.1% accuracy, and the maximum AUC measure was 0.942. Gradient Boosting (GB) showed the best performance in both methodologies followed by Support Vector Machine (SVM) for raw data and Gaussian Naive Bayes (NB) for extracted features.

Considering the size of the data and small number of participants, these results show the potential of using existing data to build more accurate and versatile HAR models.

Furthermore, ML algorithms were trained using both raw data and manually extracted features. Extracted features helped us to decrease the computational cost by up to 91.6%. However, in most cases, this reduction was achieved by undermining classification performance. The main observation was the increase in computation cost of the K-Nearest Neighbors Vote (KNN) model, which was more than 10-fold. Considering nearly 12% reduction in accuracy, the feature extraction is not justifiable for this model. Similarly, considering a 10.3% increase in training time and an 8.8% decrease in classification accuracy, using manually extracted features in SVM models is not a suitable technique for this application.

Applying principal component analysis seems a promising solution for future work to reduce the dimensionality of the dataset's features through intelligent algorithms, which may outperform manual data engineering. Artificial Neural Networks (ANN) and Deep Learning (DL) can be used to extract these features and improve human activity recognition. These methods can be applied to the proposed framework integrated into biomechanical model to improve the ergonomics [41] and human comfort [42][43].

## APPENDIX

TABLE A.1 shows optimal hyperparameters determined by Grid Search method for this study.

TABLE A.1 OPTIMISED HYPERPARAMETERS FOR TRAINING ML ALGORITHMS OVER RAW DATA AND EXTRACTED FEATURES.

| Classifier | Hyperparameter | Raw Data | Extracted Features |
|---|---|---|---|
| GaussianNB | var_smoothing | 0.000187382 | 0.0004328761 |
| DecisionTree | max_features | log2 | auto |
|  | min_samples_leaf | 2 | 5 |
|  | min_samples_split | 10 | 13 |
|  | random_state | 100 | 100 |
| Random Forest | criterion | gini | gini |
|  | min_samples_leaf | 2 | 1 |
|  | min_samples_split | 9 | 5 |
|  | n_estimators | 16 | 8 |
|  | random_state | 100 | 123 |
| KNN | algorithm | auto | auto |
|  | leaf_size | 1 | 1 |
|  | n_neighbors | 6 | 6 |
|  | weights | uniform | uniform |
| Gradient Boosting | learning_rate | 0.01 | 0.01 |
|  | max_depth | 7 | 6 |
|  | n_estimators | 50 | 50 |
|  | subsample | 0.5 | 0.5 |
| SVM | C | 1 | 0.0001 |
|  | gamma | 0.00001 | 1.00E-06 |
|  | kernel | rbf | linear |